**Correlation between radiation damage and magnetic properties in reactor vessel steels**

R. A. Kempf[(1)*], J. Sacanell[(2)], J. Milano[(3)], N. Guerra Méndez[(2)], E. Winkler[(3)], A. Butera[(3)], H. Troiani[(4)], M. E. Saleta[(3)], A. M. Fortis[(5)]

(1) División Caracterización, GCCN, CAC-CNEA, Argentina.
(2) Departamento Física de la Materia Condensada, GIyA, CAC-CNEA, CONICET, Argentina.
(3) División Resonancias Magnéticas, CAB-CNEA, CONICET, Argentina.
(4) División Física de Metales, CAB-CNEA e Inst. Balseiro (UNCU), CONICET, Argentina.
(5) Departamento Estructura y Comportamiento. Gerencia Materiales-GAEN, CAC-CNEA, Argentina

* e-mail  kempf@cnea.gov.ar

**Abstract**

Since Reactor Pressure Vessel steels are ferromagnetic, provide a convenient means to monitor changes in the mechanical properties of the material upon irradiation with high energy particles, by measuring their magnetic properties. Here, we discuss the correlation between mechanical and magnetic properties and microstructure, by studying the flux effect on the nuclear pressure vessel steel used in reactors currently under construction in Argentina. Charpy-V notched specimens of this steel were irradiated in the RA1 experimental reactor at 275°C with two lead factors (LFs), 93 and 183. The magnetic properties were studied by means of DC magnetometry and ferromagnetic resonance. The results show that the coercive field and magnetic anisotropy spatial distribution are sensitive to the LF and can be explained by taking into account the evolution of the microstructure with this parameter. The saturation magnetization shows a dominant dependence on the accumulated damage.

Consequently, the mentioned techniques are suitable to estimate the degradation of the reactor vessel steel.

# 1. Introduction

The integrity of the components of power plant reactors throughout their service life is affected by the degradation suffered by the materials. This degradation arises from unfavorable environmental conditions, mainly due to the action of radiation [1].

From the viewpoint of nuclear technology, the fundamental importance of the embrittlement phenomenon of a pressure vessel is that, being a non-redundant or replaceable component of the primary system in a nuclear power station, the plant lifetime is limited to a period in which the container properties are reliable.

The reactor pressure vessel steel (RPV), composed mainly of iron, and is naturally ferromagnetic. This offers the possibility to evaluate the mechanical performance of the material and then relate it to its magnetic properties. Due to the high Curie temperature of iron ($T_c = 770$ °C), ferromagnetism persists under normal operating circumstances in the reactor. The development of a characterization technique that relates mechanical and magnetic properties is desirable because it can result in an *in situ* on-line monitoring system for reactor pressure vessels once the correlations have been established. Some studies have focused on the connection between microstructural factors and magnetic responses [2,3,4,5]. However, the effect of radiation damage on magnetic properties has received little attention. The work presented here is part of the study of the dose-rate effect on RPV embrittlement. In a previous work, we examined the flux effect on the low fluence regime by studying irradiated A 508 class 3 steel at the RA1 experimental reactor [6]. In that work, we observed a shift of the ductile-brittle transition temperature (DBTT) to larger values for the sample irradiated for a longer time.

To ensure the plant's integrity, monitoring programs are carried out by irradiating steel samples in an accelerated form to follow the evolution of the material at different irradiation doses. In these monitoring programs, Charpy tests are performed on irradiated samples in certain places of the reactors under surveillance that receive a larger neutron flux than the container itself, or on samples irradiated in experimental reactors. However, due to the nature of radiation damage, the results obtained in accelerated irradiations are not always useful to be compared with what really happens to the container in service.

A shift in DBTT occurs in a ferritic steel sample upon irradiation with neutrons. This can be verified experimentally by analyzing the results of Charpy impact tests. This technique measures the energy absorbed by the sample during the impact, before and after irradiation. An increase in the DBTT means that the container could be in conditions of fragility in cases in which the temperature abruptly decreases due to an accident (loss of coolant accident, LOCA).

Studies based on thermodynamics, kinetics and micromechanics of the damage process induced by neutron irradiation of materials propose two mechanisms for embrittlement [7, 8]:

1) Matrix damage due to the formation of aggregates of point defects and dislocation loops during irradiation (stable matrix damage, SMD).
2) Formation of precipitates rich in copper and other alloying elements, which are favored by neutron irradiation (Cu-rich precipitates, CRP).

Both processes result in an increase in the number of obstacles for dislocation movement, which in turn causes embrittlement.

SMD is the dominant damage process for low-Cu steels. The resulting dispersed barrier irradiation hardening increases roughly with the square root of the fast fluence and decreases as the irradiation temperature increases [9]. On the other hand, CRP damage is linked to irradiation-enhanced formation of Cu-enriched solute clusters. Copper-rich clusters or precipitates are formed as a residue of the annealing of vacancy-copper complexes or by normal nucleation due to the high copper super-saturations. The contribution of CRP damage to the total DBTT depends on the Cu and Ni content of the steel and on the strong Cu-Ni interaction [10]. The CRP damage is a diffusion process where the time, together with the rate of point-defect production, plays an important role.

Our group has previously performed irradiation experiments on pressure vessel steels of nuclear reactors currently under construction in Argentina (CAREM and Atucha II) with different accelerations, in order to determine the effect of irradiation on their mechanical behavior, including their fragility. Those accelerations are characterized by the lead factor (LF). LF is the ratio between the instantaneous neutron flux density in the monitoring specimen position ($\phi_M$) and the maximum neutron flux density calculated on the inner surface of the container wall of the reactor pressure vessel ($\phi_{RPV}$). It gives a measure of the acceleration of irradiation:

LF=$\phi_M$ (E > 1MeV) /$\phi_{RPV}$ (E > 1MeV).

Ferritic steels that change from ductile to brittle can fracture when their deformation temperature falls below a critical value (the DBTT). Neutron irradiation increases the yield stress and decreases ductility. So, irradiation embrittlement is typically characterized by an increase in the DBTT, marking the transition between the low toughness cleavage and high toughness ductile fracture regimes. Taking into account the dispersion presented by Charpy tests [6], here we continue our study focusing on the magnetic properties, which have been recently proposed as a complementary technique to be considered in an *in situ* monitoring program [11].

Previous studies of the correlation between dislocations and magnetic properties have shown that the strain field around dislocations influences the magnetization hysteresis loops through magnetoelastic coupling in distinct ferromagnetic materials [12].

Park et al. analyzed the degradation induced by neutron irradiation in ferritic steels with larger fluence than that used in the present study [13]. These authors measured ferromagnetic resonance (FMR) and DC magnetization in order to characterize the damage produced in the material. They observed a slight decrease in the saturation of the magnetization ($M_{sat}$) and an increase in the FMR linewidth, which suggests that both techniques can be applied in surveillance. The relation between those magnetic changes and the micromechanisms involved in the embrittlement process has been studied in [4], [5] and [11]. However, in order to enhance the quantities measured, the authors selected another kind of steel with a large amount of alloying elements.

In the present work, the microstructural characterization of the damage and the distribution of copper and manganese-rich precipitates were explored by transmission electron microscopy (TEM). FMR and magnetic hysteresis loops were measured to deal with the changes in the magnetic properties due to irradiation. The selected methods require a very small quantity of sample, which also ensures the possibility of safe handling by the researchers.

Magnetization measurements consisted of performing hysteresis cycles in order to observe the change in shape of these curves and obtain the coercive field $H_c$ and the saturation magnetization $M_{sat}$. The coercive field is extremely sensitive to the presence of defects in a ferromagnetic material. For example, it increases as the density of defects increases and also when internal stresses increase.

We studied the changes in these parameters after neutron irradiation, which generates the agglomeration of defects and favors the creation of nanoprecipitates. On the other hand, FMR measurements showed signatures of a structural inhomogeneity evidenced by a broadened distribution of magnetic anisotropies.

There are precedents in the literature regarding the study of the magnetic properties of irradiated materials which show that the creation of Cu-Ni-Mn-rich nanoprecipitates in RPV steels is favored by neutron irradiation [4,5,8].

In this work, we propose a systematic study of the magnetic properties complemented with the corresponding microstructural analysis. We show that the presence of nanoprecipitates causes changes in both $H_c$ and $M_{sat}$, suggesting that they can be easily used to characterize damage in RPV steels.

## 2. Materials and Methods

Irradiations were performed to determine the effect of different LFs on the mechanical behavior of ASME SA-508 class 3 steels (see Table I), which will be used in Atucha II [6]. The temperature during the measurements was set at power reactor conditions (T= 275 ºC) and the experiments were carried out in an experimental device specially designed for this purpose. Samples were irradiated keeping the integrated dose constant, but with different times and LFs for the same fluence.

Table I – Steel´s composition (%Wt)

| C | Mn | P | S | Si | Ni | Cr | Mo | V | Cu | Al | Ta |
|---|---|---|---|---|---|---|---|---|---|---|---|
| 0.25 | 1.40 | 0.012 | 0.015 | 0.15 | 0.74 | 0.20 | 0.53 | 0.03 | 0.10 | 0.05. | 0.03. |

Neutron flux in the inner wall of the RPV of Atucha II (PHWR, CAN II) was estimated as $\phi$ (E>1 MeV) = $2\times10^{13}$ nm$^{-2}$s$^{-1}$. Then, for 40 years calendar of life by design, the fluence is about $2.5\times10^{22}$ nm$^{-2}$. Two sets of irradiations were performed at a fluence of $6.6\times10^{21}$ nm$^{-2}$, which corresponds to about 10 years of operation of the CNAII (see Table II).

Table II –Irradiation plan

|  | Flux[n/m$^2$s] | Fluence[n/m$^2$] | LF | IrradiationTime (hs) | T [°C] |
|---|---|---|---|---|---|
| First set | 3.715X10$^{15}$ | 6.6X10$^{21}$ | 186 | 492 | 275 |

| Second set | 1.857X10$^{15}$ | 6.6X10$^{21}$ | 93 | 984 | 275 |

After the irradiation and the Charpy assays, discs of 3 mm in diameter and 100 micrometers thick were extracted from each sample. These discs were further used to measure magnetization, FMR and for TEM. Table II shows the irradiation parameters. Magnetization measurements were performed in a Versalab$^{TM}$ (Quantum Design) vibrating sample magnetometer (VSM), in the range from 50 to 400 K and magnetic fields up to 3 Tesla. The FMR study was performed in a Bruker electronic spin resonance spectrometer (ESR300), at an operating frequency of 9.4 GHz (X band) with an applied magnetic field up to 1.8 Tesla. Angular scans were performed at room temperature by rotating the field from the disc plane to the disc normal. TEM images were obtained with a Philips CM 200 (Ultra Twin lens and LaB6 filament) operated at 200 kV.

**3. Results and Discussion.**

Figure 1 shows the TEM images for the non-irradiated sample (a) and for the samples with LF=93 (b) and LF=186 (c). The presence of precipitates is evident in the irradiated samples (Fig. 1 (b) and (c)), which show a larger precipitate size for the LF=93 samples. This difference is a clear sign of a nucleation and growth process during irradiation, due to the longer irradiation time of the LF=93 samples with respect to the LF=186 ones. The mean diameter ($D$) and mean distance between them ($D_p$) are summarized in table III.

Table II –Irradiation plan

|  | Flux[n/m$^2$s] | Fluence[n/m$^2$] | LF | IrradiationTime (hs) | T [°C] |
|---|---|---|---|---|---|
| First set | 3.715X10$^{15}$ | 6.6X10$^{21}$ | 186 | 492 | 275 |
| Second set | 1.857X10$^{15}$ | 6.6X10$^{21}$ | 93 | 984 | 275 |

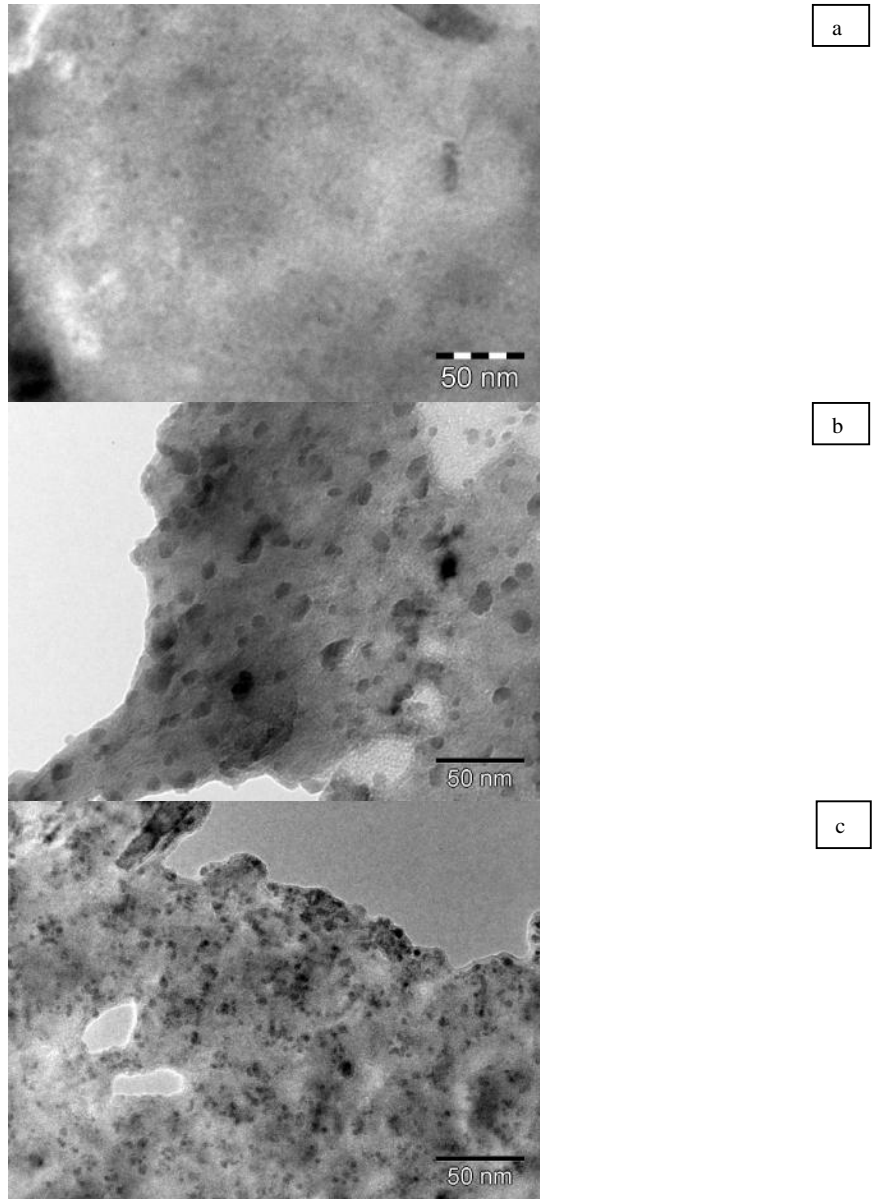

Figure 1. TEM micrographs for a) non irradiated, b) LF 93 and c) LF 183 samples.

The sample irradiated at LF=186 showed precipitates of $D = 5$ nm, separated by $D_p \sim 7$ nm, while the sample irradiated at LF=93 showed $D \sim 8.5$ nm, separated by $D_p \sim 25$ nm. The occurrence of the observed distribution of defects is responsible for the embrittlement of steels under irradiation. According to this, the most recent models of RPV embrittlement [14] include a component of hardening by dispersion based on the model of ageing developed by Russell and Brown [15], which are modified to take into account the diffusion rate increase caused by the over-saturation of vacancies that results from radiation. As it can be clearly seen, this distribution strongly depends on the LF, even for the same neutron fluence. The resulting morphology, showing smaller defects and larger spatial density for the LF=186 sample than for the LF=93 one, is expected to affect any physical property sensitive to local variations in the material morphology, such as the magnetic domain wall displacement that takes place when a ferromagnetic material is magnetized. In that sense, a larger tortuosity is expected in the sample with small defects that are close to each other (i.e. LF=186 sample) [16].

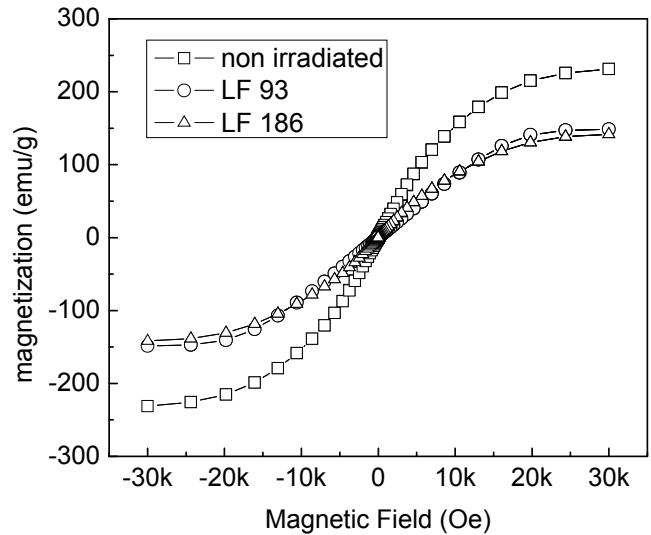

Figure 2. Magnetization hysteresis loops. A reduction of the saturation of the magnetization is observed for the irradiated samples.

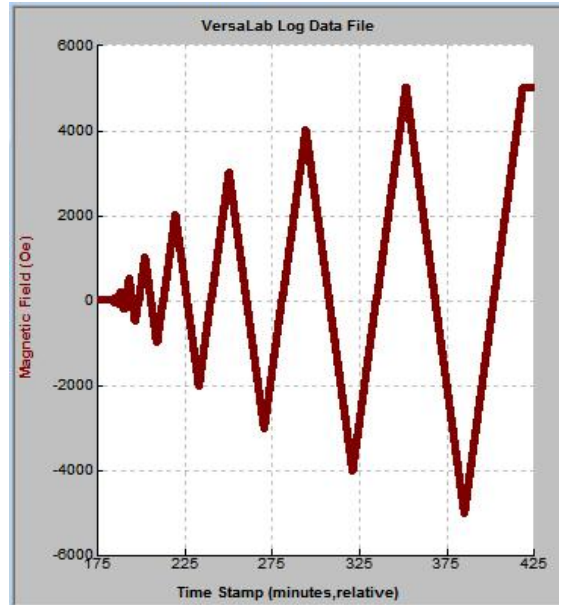

Figure 3: Schematic of the magnetic field protocol in the minor loops (see text).

Figure 2 shows the magnetic hysteresis loops for the three samples (non-irradiated, LF=93 and LF=186). Measurements were performed with the magnetic field in the plane of the discs. When comparing the non-irradiated sample with both irradiated ones, $M_{sat}$ clearly decreases by about 30% upon neutron irradiation. A less significant decrease is observed between the LF=93 and the LF=186 samples. This feature seems to indicate that the reduction of $M_{sat}$ is related with the SMD mechanism due to the appearance of agglomerates of point defects and dislocation loops during the irradiation .However, it is not clear that other mechanisms could be contributing to the observed change.

As mentioned before, the distribution of nanoprecipitates is expected to hinder the movement of magnetic domain walls by introducing barriers on their path throughout the material. Also, due to the nucleation and growth process that occurs during irradiation, different LFs result (as shown in figure 1) in a different distribution of nanoprecipitates. The influence of those distributions on the magnetic properties was studied through their effect on the coercive field ($H_c$), which is related to the tortuosity of the path for the domain walls.

However, we have to take into account that a large enough magnetic field will serve to pass every possible barrier, hiding the influence of the nanoprecipitates generated by irradiation, and that a magnetic field threshold above which the effect of irradiation will not be observed should thus exist. Then, in order to study differences between different LFs, minor loops were also studied [4, 11].

The experiment consisted in performing different magnetic loops varying the magnetic field from 0 to $H_{max}$, to $-H_{max}$ and then again to 0 to close the cycle. This procedure has to be made for several $H_{max} < H_{sat}$, being $H_{sat}$ the field needed to reach $M_{sat}$. Figure 3 schematically shows the H vs time protocol used. We extracted $H_c$ from the data as follows: we took $H_c^+$ ($H_c^-$) as the magnetic field needed to return to zero magnetization after reaching $H_{max}$ ($-H_{max}$). Then, we saved $H_c$ as the mean value of $|H_c^+|$ and $|H_c^-|$.

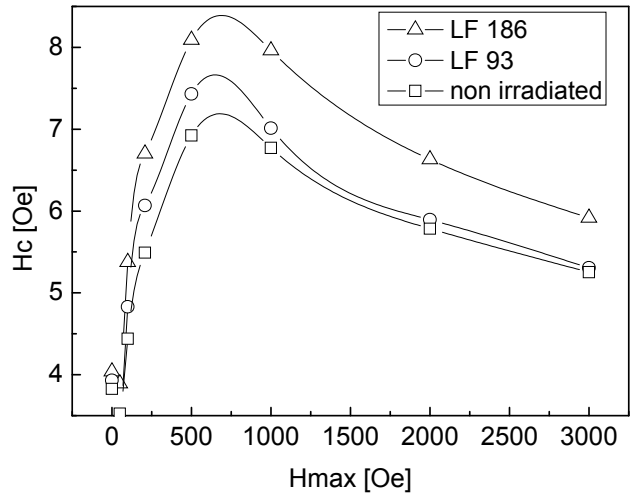

Figure 4: Coercive field as a function of $H_{max}$ in the minor loops (see text).

The $H_c$ vs. $H_{max}$ dependence is shown in figure 4. A first look shows an enhanced value of $H_c$ for the irradiated samples, in accordance with our hypothesis.

For $H_{max}$ larger than 2000 Oe, the difference between the non-irradiated sample and the one with LF=93 is virtually absent, showing that the threshold mentioned before was reached.

The fact that $H_c$ for the sample with LF=93 is smaller than that corresponding to the sample with LF=186 can be explained by comparing the $H_c$ results with the TEM images from figure 1. In the case of LF=93, the presence of large precipitates separated by a comparatively long distance between them leaves enough space for the passage of the domain walls, while the smaller precipitates for the LF=186 sample, which are closer to each other, form an almost continuous barrier that hinders the movement of the domain wall. An intriguing behavior was observed in the trend of all measurements for which the explanation is not clear at present. $H_c$ shows a maximum value for $H_{max} \sim 500$ Oe and then decreases if we further increase $H_{max}$. Further studies are needed to elucidate this point.

The changes in $M_{sat}$ and $H_c$ from DC magnetometry have shown to be highly influenced by the spatial distribution of the precipitates [16]. A deeper understanding can be achieved by analyzing how the inhomogeneity created by that distribution can affect domain wall movement, this study that can be made through FMR.

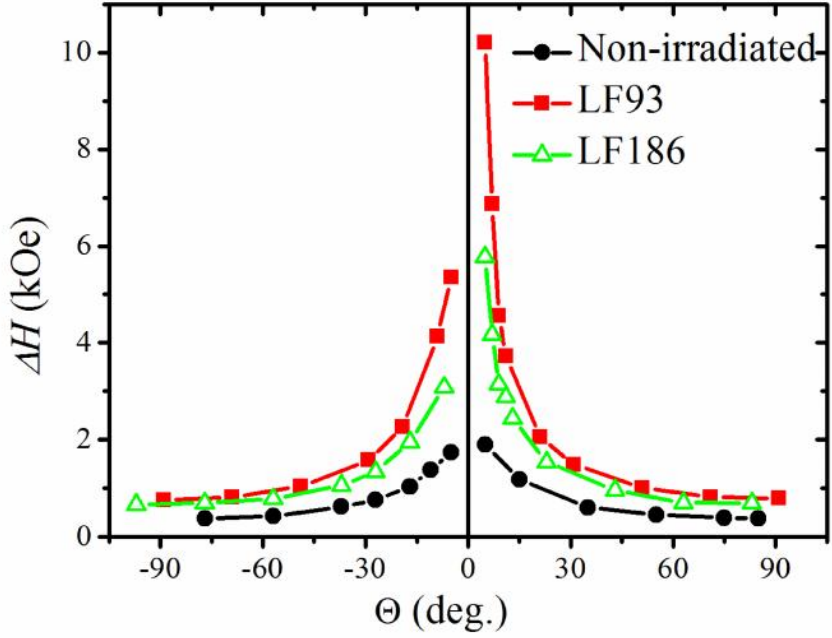

Figure 5: FMR linewidth, ∆H, as a function of the polar angle, Θ, for the all studied samples.

H. Kronmuller et al. presented a model that shows that $H_c$ increases linearly with the density of defects and decreases as the size of precipitates increases [17]. Then, the increase in the density of defects associated with the SMD component allows us to qualitatively explain the increase in $H_c$ for the irradiated samples, which have a larger density of defects than the non-irradiated ones. Additionally, for the irradiated samples, $H_c$ decreases as the size of the precipitates increases, i.e. the sample with LF=93 has a lower $H_c$ value than the one obtained for the sample with LF=186, a result which is also in accordance with the model presented by H. Kronmuller et al. In spite of showing a behavior that matches with the Kronmuller's model, the changes observed in $H_c$ are small and seem to be reduced for another effect. In this sense, it should be taken into account the domain wall width. The Kronmuller's model assumes that the domain wall is much wider than the precipitate average diameter, but it is not the case for our irradiated samples. For pure iron the wall width is around 20 nm while the precipitate diameters are around 5 nm and 8 nm. This fact produces that the obstacle role plays by the precipitates is diminished, i.e., the resistance to the wall movement will be an average of that sensed along the wall width.

Figure 5 shows the linewidth, ΔH, of the FMR spectra for the samples studied, as a function of the angle between the magnetic field externally applied, $H_{ext}$, and the sample surface. The curves show the typical behavior for ferromagnetic thick films like our discs. The linewidth increases when the angle approaches 90º (the film's plane perpendicular to the applied field) as expected for samples with a large demagnetizing factor. To understand this, we briefly summarize the physical principles of FMR. In thick films, in the remnant state, the magnetization lies in the film's plane in order to reduce the magnetostatic energy. If a magnetic field is applied out of the film's plane, the magnetization of the sample will tend to be aligned to it. Then, the equilibrium angle for the magnetization vector will be the result of the competition between the strength and the angle of the applied field, the demagnetizing factor and $M_{sat}$ [18]. The equilibrium conditions for the magnetization enter as an input to obtain the resonance field, $H_r$. For our samples, the values of the linewidth observed are related to a broadening of the spectra due to the magnetic inhomogeneities present in the samples. Then, we measured a continuous range of $H_r$. Moreover; the effects due to the inhomogeneities are enhanced when $H_{ext}$ approaches 0º and the linewidth become larger, as observed in the experiments.

As mentioned above, the three samples show the same behavior as a function of the polar angle. However, ΔH for a given angle is different for the three samples, i.e., ΔH for the non-irradiated sample is always lower than that for the irradiated ones. Among the irradiated samples, the LF=93 is the one with higher ΔH. For instance, if we take

$_H$=90º, ΔH (non-irradiated) = 0.39 kOe, ΔH (LF=186) = 0.72 kOe, ΔH (LF=93) = 0.82 kOe. As stated in the previous paragraph, ΔH arises mainly from the inhomogeneities that induce local variations in the magnetization throughout the sample. Within this context, the results for the non-irradiated sample are consistent with the fact that its structure is more homogeneous than that of the irradiated ones. Also, for small LFs, the inhomogeneities effect seems to be increased, also in agreement with our TEM images. In this sense, we can argue that neutron irradiation causes local changes in the magnetization in the sample, which are related to microstructure variations. In TEM micrographs, we can observe how the aggregate size is increased for longer exposure times (decreasing LFs). Larger aggregates lead to a more inhomogeneous behavior and then larger local magnetization changes. This can be understood if we take into account the role of the magnetic parameter called exchange length ($l_e$). This parameter gives a maximum distance in which a change in the magnetic state is sensed.

Focusing in our sample, $l_e$ gives the distance around a single aggregate, where the magnetization value is perturbed due to the presence of such aggregate. In particular, the precipitates are the main reason for the magnetization change. In ferromagnetic materials, like the one studied in the present work, $l_e$ is around 2 to 3 nm [15]. If we take into account the mean distance among precipitates, it is possible to relate the TEM micrographs with the linewidth measured by FMR. The absence in the non-irradiated sample (Figure 1) implies that the magnetization is homogeneous and, as expected, the FMR linewidth is the smallest of all the samples. When the irradiation process is set, the precipitates create regions where the magnetization is depressed. As mentioned above, the amount of material around the precipitates that senses the magnetization change is related to $l_e$. From the micrograph, it is possible to determine that the mean distance among the precipitates is 25 nm for the sample with LF = 93 and 7 nm for the sample with LF = 186. On the one hand, for the LF = 93 sample, $D_p$ is much larger than $l_e$. This leads the material to present well-distinguished regions with different magnetization, i.e., the regions at distances to the precipitates smaller than $l_e$ and those larger than $e_l$. On the other hand, for the sample with LF = 186, $D_p$ is the same order as $e_l$. This causes the whole ferromagnetic material to behave as a single precipitate. In this sense, there are no well-determined regions with different local magnetization and, in average, the magnetization fluctuations along the sample are reduced with respect to those of the LF = 93 sample. Therefore, the absorption signal expected for the LF = 93 sample should be broader than that measured for the LF = 186 one, as observed in the experiments.

## 4. Conclusions.

The purpose of this work was to correlate changes in the magnetic properties of nuclear reactor pressure vessel steels with the progression of changes in the mechanical properties induced by irradiation. Magnetic properties were measured after samples were irradiated with two lead factors, giving different results as a consequence of the changes in the structure of the radiation damage that occurs in different irradiation times.

TEM studies were used to observe the nucleation and growth process of the precipitates, showing an increased size and separation for longer irradiation times (at an equal overall dose). The analysis of the TEM images shows that irradiation causes an environment for precipitates, probably Cu-Ni-Mn ones, whose size and distribution determine the hardening and embrittlement of the steel. The appearance of the spatial inhomogeneity generated by the precipitates is put in evidence magnetically through FMR which shows an increasing anisotropy distribution with such inhomogeneity.

Te most significant variation observed in the magnetic measurements was the almost 30% decrease in $M_{sat}$ of the irradiated samples, an easier magnitude to measure that seems to be directly related to the materials microstructure.

The investigation through magnetic hysteresis minor loops showed that that the effect evidenced in the TEM images can be inferred using this simple and non-destructive method, which can be used as an alternative way to monitor the steel's mechanical properties. However, further studies are needed in that direction in order to find a direct relation between the embrittlement, the changes in the size and distribution of the precipitates and the consequence of this relation on the coercive field.

A next step that arises from this investigation would be the identification of the defects and their distribution. This would provide additional information to characterize the relation between embrittlement and the changes in the magnetic properties. The goal will be to develop an *in situ* method to monitor the mechanical state of RPV steels in operation. Our group is currently working in that direction.

**Acknowledgments**

This work was supported by the IAEA Technical Co-operation Project ARG/4/093-92 04 (SUPPORTING AGEING MANAGEMENT FOR THE LONG-TERM OPERATION OF ATUCHA II NUCLEAR POWER PLANT) and by CONICET (PIP 00038/2008).

We thank Daniel Anello for the support in the experimental design of the irradiation devices.